\documentclass[11pt,twoside]{article}
\usepackage{asp2004}
\usepackage{psfig}
\usepackage{epsf}
\usepackage{graphics}
\usepackage{lscape}
\def\arg#1{{\it#1\/}}
\let\prog=\arg
\def\edcomment#1{\iffalse\marginpar{\raggedright\sl#1\/}\else\relax\fi}
\reversemarginpar

\begin{document}
\title{On the Origin of Sulfur}
 \author{Nils Ryde}
\affil{Department of Astronomy and Space Physics, Uppsala University, SE-751 20 Uppsala, Sweden\\ \texttt{ryde@astro.uu.se}}
 \author{David L. Lambert}
\affil{The W.J. McDonald Observatory,  University of Texas at Austin, Austin TX 78712, USA\\ \texttt{dll@astro.as.utexas.edu}
}

\begin{abstract}
We present our work on the halo evolution of sulfur, based on
observations of the S I lines around
$9220$~\AA\ for ten stars for which the S
abundance was obtained previously from much weaker S I
lines at $8694$~\AA .
We cannot confirm the rise and the high [S/Fe] abundances for low [Fe/H],
as  claimed in the literature from analysis of the $8694$~\AA\ lines.
The reasons for claims of an increase in [S/Fe] with
decreasing [Fe/H] are probably twofold: uncertainties in the measurements of the
weak 8694 \AA\
lines, and systematic errors in metallicity determinations from
Fe I lines.
The near-infrared sulfur triplet at ~$9212.9$, $9228.1$, and  $9237.5$\,\AA\
are preferred for an abundance analysis of sulfur for metal-poor stars.
Our work was  presented in full by \citet{ryde}.
\end{abstract}

\begin{figure}
\label{fig1}
\caption{Part of the Grotrian diagram of sulfur is shown with a few transitions indicated.
The transitions in black indicate the IR triplet lines at 9212.9, 9228.1, and
9237.5 \AA\ and the weaker 8694 \AA\ lines used by \cite{israel} and \citet{takeda}. The Figure
is based on the Grotrian diagrams in \cite{term}. }
\plotone{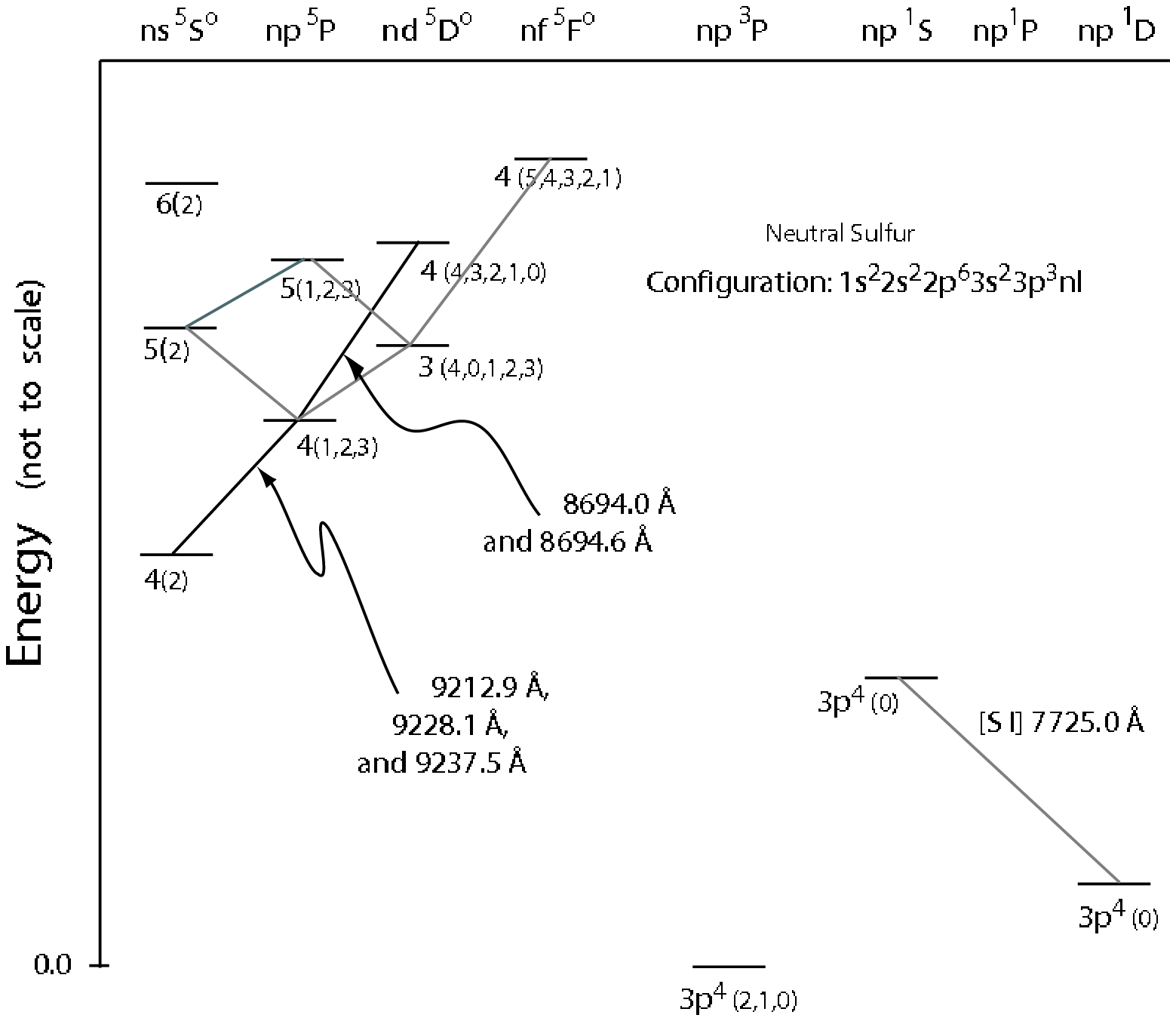}
\end{figure}

\begin{figure}
\label{fig2}
\caption{The chemical evolution of sulfur. Our new measurements are shown by star symbols.
These are connected to the earlier observations, claiming a rise. The measurements of \citet{nissen_s} and
\citet{chen} are also shown.}
\plotone{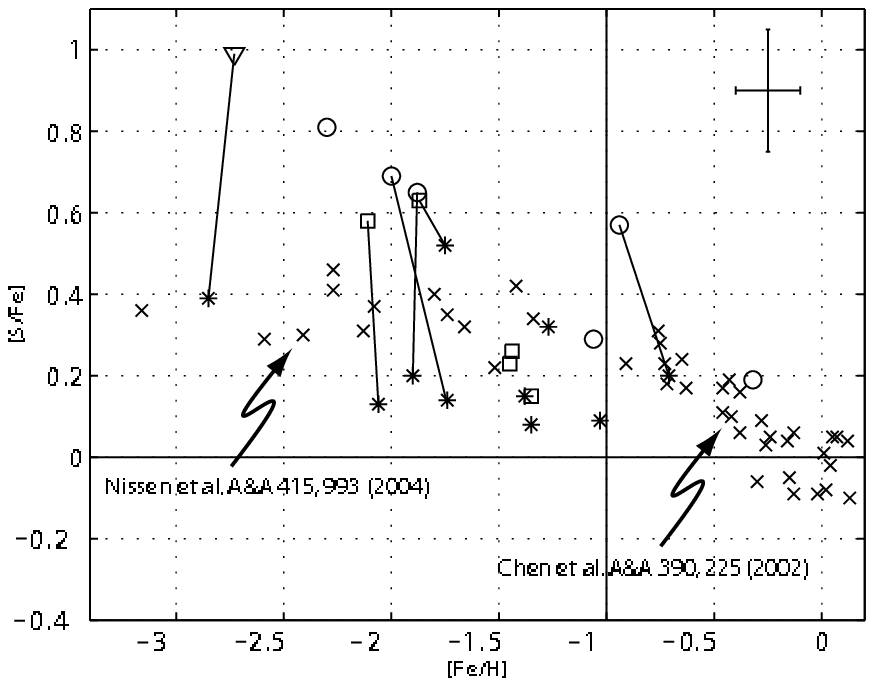}
\end{figure}

\begin{figure}
\caption{Example of our observations. The telluric lines in the spectrum above are nicely
eliminated in the final one below.}
\plotone{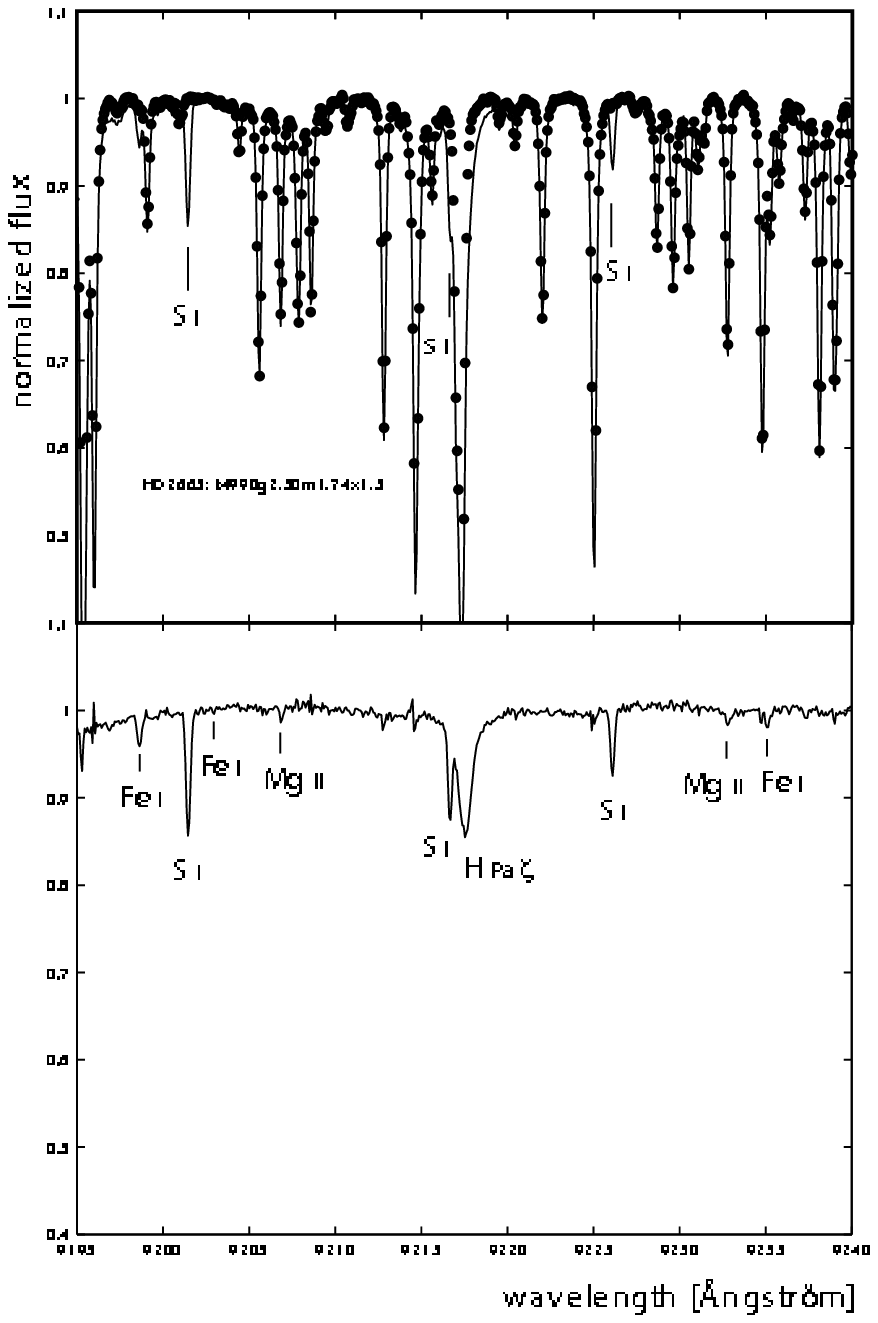}
\end{figure}



There has recently been a
debate in the literature on the chemical evolution of sulfur in the
halo phase  of the Milky Way ([Fe/H] $\leq -1$).
\cite{israel} and \citet{takeda} claim a monotonic
rise in [S/Fe] for decreasing [Fe/H] based on an analysis using weak sulfur
lines at 8694 \AA\ measured in spectra of approximately ten stars each.
 This rise has consequences for our understanding of the sites
of formation of sulfur in the
Universe.

Is there a rise or not?
In order to answer this question, we have reinvestigated ten of the stars examined
previously, by analysing
the infrared triplet lines of neutral sulfur at 9220 \AA\ instead of the weaker 8694 \AA\
lines used before. The lines are indicated in the Grotrian diagram shown in Figure 1.
The lower levels of the 8694  \AA\ lines are the upper levels of the 9220  \AA\ lines
and hence the excitation potential difference is 1.34 eV.


We observed the stars with the 2dCoud\'e spectrometer (Tull et al. 1995)
 at McDonald
Observatory in 2001. Our stars are a mix of
 dwarfs and giants spanning a temperatures range of $4200-6300$ K, and a
metallicity range from [Fe/H]$=-0.7$ down to [Fe/H]$=-2.9$. Exposure
 times ranged from
half-an-hour to three hours per star.
In our analysis, we use the  model atmosphere parameters
 used by \cite{israel} and \citet{takeda} with
the exception that the metallicity is re-determined from Fe II lines, which are not much
affected by non-LTE effects, instead of the more
uncertain determination of the metallicity based on an non-LTE analysis of Fe I lines.

We do not confirm the rise in [S/Fe] with decreasing
[Fe/H] in the halo phase.
 Instead, we confirm  that [S/Fe] attains a plateau for
the halo phase of the Galaxy, also shown by \citet{nissen_s}, see Figure 2.
Thus, sulfur  behaves similarly to alpha elements (e.g., Mg, Si, and Ca), indicating that the
primary source for sulfur atoms in the Universe is Supernovae Type II explosions.

A significate advantage of the near-IR lines over the 8694~\AA\
 lines in the study of halo stars is their factor
of ten larger equivalent widths.
An apparent disadvantage of the near-IR lines is the ubiquitous telluric vater-vapor
lines in this region.
By observing a  rapidly-rotating hot star at a similar air mass to each
halo star observation, the water vapor lines may be divided out, as
shown in Figure 3.

\end{document}